# Dynamics and Energy Distribution of Non-Equilibrium Quasiparticles in Superconducting Tunnel Junctions

K. Segall, C. Wilson, L. Li, L. Frunzio, S. Friedrich, M.C. Gaidis and D.E. Prober

**Abstract:** We present a full theoretical and experimental study of the dynamics and energy distribution of non-equilibrium quasiparticles in superconducting tunnel junctions (STJs). STJs are often used for single-photon spectrometers, where the numbers of quasiparticles excited by a photon provide a measure of the photon energy. The magnitude and fluctuations of the signal current in STJ detectors are in large part determined by the quasiparticle dynamics and energy distribution during the detection process. We use this as motivation to study the transport and energy distribution of non-equilibrium quasiparticles excited by x-ray photons in a lateral, imaging junction configuration. We present a full numerical model for the tunneling current of the major physical processes which determine the signal. We find that a diffusion framework models the quasiparticle dynamics well and that excited quasiparticles do not equilibrate to the lattice temperature during the timescales for tunneling. We extract physical timescales from the measured data, make comparisons with existing theories, and comment on implications for superconducting mesoscopic systems and single-photon detectors.

## I. Introduction

Since Giaever's first experiments[1] on electron tunneling in superconducting tunnel junctions, it has been well known that tunneling is an excellent probe of both electron energy distribution and density of states. Many years of experiments have followed in which superconducting tunneling structures at or near equilibrium have shown interesting phenomena.[2] More recently, superconducting tunnel junctions have been used[3] effectively as single photon spectrometers, for photons in the energy range 1-$10^4$ eV. Photons absorbed in a superconducting tunnel junction create quasiparticle excitations, and by measuring the resulting increase in tunneling current the photon's energy can be determined. In our devices the photons are absorbed in a tantalum (Ta) film, and tunnel through aluminum-based (Al) tunnel junctions. The physics of the charge collection and readout in these detector structures follows directly from much of the earlier work on tunneling. However, two factors make the analysis of these devices more difficult. The first (1) is that due to the need for large absorbing films, the quasiparticle transport to the tunnel barrier is not instantaneous, resulting in complex dynamics. The second (2) is that photon induced quasiparticles do not equilibrate to the lattice temperature on a timescale over which the tunneling takes place, making it a non-equilibrium situation. The result of these two factors is that the measured tunneling current (signal) and its fluctuations (noise) are often difficult to explain theoretically.

In three recent papers we explored some of these issues with devices fabricated and measured in our group. In reference 4, we presented some measured timescales for various quasiparticle processes in an effort to address (1) above.[4] We studied the effects of diffusion, trapping, tunneling, recombination and inelastic scattering on the dynamics of the current pulse. In reference 5, we looked in more detail at the loss and diffusion in Ta films, and also discussed how the measured timescales are affected by the absorber length.[5] Reference 6 examined the effect of a heated quasiparticle energy distribution on the signal fluctuations, leading to insights into (2) above.[6] We found that the elevated effective temperature of the quasiparticles resulted in new noise sources, helping to explain the measured energy resolution.

In this report we present a full theoretical and experimental study of our aluminum superconducting tunneling structures, in an effort to more fully elucidate the

--

physics behind the difficulties presented by (1) and (2) above. We derive a full numerical model of the tunneling current. We account for the spatial dynamics through a diffusion calculation and the energy distribution through an iterated set of rate equations. The calculated tunneling currents from this model are compared directly to the experimental tunneling current, with excellent agreement. The results from this model have been used in previous work,[4-6] but the model itself has yet to be presented in detail until now. The agreement between the model and experiment validates the use of a diffusion framework to describe the spatial dynamics and yields fitted values for many of the junction timescales and physical parameters. The dependence of the tunneling current on the junction voltage and circuit impedance is also fit with the model. These dependences show that the quasiparticle energy distribution is at an elevated effective temperature, reducing the effective junction impedance. We detail the effects of this heated distribution on the tunneling current. Finally, we show the impact of differing values of the electron-phonon scattering rate in our Al films from the dependence of the tunneling current on temperature and voltage, arriving at a rough estimate for the value of this important parameter.

These results *on the non-equilibrium dynamics* have implications for experiments on mesoscopic and superconducting systems and for the treatment of quasiparticle tunneling and dynamics out of equilibrium. Our work shows how to treat such a time-dependent non-equilibrium system. We find that characterizing an effective temperature of the quasiparticle system is a reasonably good description on the microsecond *tunneling* timescale of our experiment. This work also has important consequences for the performance and design of future superconducting detectors, and possibly for superconducting quantum-computation circuits.

Earlier work in this field treated the lateral diffusion of quasiparticles in imaging-type detector geometries, but did not address the energy distribution during tunneling.[7] More recent work has treated the quasiparticle energy distribution in detail, including the effects of recombination, tunneling and phonon exchange.[8] That kind of treatment is ideal for vertical trapping devices which have significant backtunneling, as the long effective tunneling time allows these processes to proceed, and makes their effects significant. Our devices, which use lateral trapping, are best treated with a model that includes *both* the effects of lateral trapping and the quasiparticle energy distribution in the trap electrode.  This model is designed to have the minimum number of fitting parameters, each of which can be constrained by experimental data.  Our model focuses on only the processes that most significantly affect the tunneling current.  It gives predictions which can be compared directly to the time-dependent current pulses. These predictions are also used to study the interaction of the junction with its external circuit impedance, an issue of importance for our class of devices.  We compare the predicted interactions to those seen in our experiments.

The paper is organized as follows.  Section II details the devices studied and the experiments performed.  In section III we review the relevant physical processes and derive the model for the tunneling current.  Section IV compares the output of the model with the data and details the effects of quasiparticle dynamics on the tunneling current. Section V looks at the temperature and voltage dependence of the tunneling current to study the quasiparticle energy distribution.  In section VI we conclude and provide outlook for future experiments.

**II. Experiments**

In this section we describe the sample geometry and fabrication, electronic readout and cryogenic testing of our tunnel junction structures.  Each of the samples studied consists of two aluminum-aluminum oxide-aluminum (Al-AlOx-Al) tunnel junctions attached to a single tantalum (Ta) absorber.  An example of such a device, with a top view and side view is shown in Fig. 1.  The use of two junctions with a single absorber gives the device inherent imaging capabilities as a photon detector; this is discussed below.  The Ta absorber is 200 μm in length, 100 μm wide and 600 nm thick. The base electrode of each tunnel junction, known as the trap, is an Al film 150 nm thick. The Al trap overlaps the Ta absorber by about 10 μm.  The tunnel barrier for each junction is aluminum oxide, with a total junction area of about 1800 μm$^2$. The junction is stretched into a quartic shape in order to reduce the value of the magnetic field required to suppress the Josephson tunneling (see below).  The counter-electrode is an Al film 80 nm thick and is covered by an Al wiring layer about 220 nm thick.  A thin strip of niobium (Nb), shown in the top view only, makes electrical contact to the absorber.  It is

--

6 μm wide, 60 μm long and 150 nm thick. The whole structure is fabricated on an oxidized silicon (Si) wafer.

All devices have been fabricated at Yale in a high-vacuum deposition system with in-situ ion beam cleaning. The Ta is put down first, sputtered at 750 C to improve the film quality. After the Ta deposition the surface is ion-beam cleaned and the Nb contact is sputtered at room temperature. The surface is again ion-beam cleaned and then, all in one vacuum cycle, the Al trilayer (Al-AlOx-Al) is deposited in two evaporations separated by one oxidation step. The tunnel barrier is oxidized to a current density of about 30 A/cm$^2$. A layer of silicon monoxide (SiO) is evaporated to passivate the junction edges and the absorber surface; this is not shown in the figure. Finally, the Al wiring layer is evaporated. An ion bean cleaning is performed prior to each metal deposition to ensure good metallic contact. All layers are patterned with photolithography, using either wet etching or lift-off. Other details of the geometry and fabrication procedure have been published.[9]

The devices were cooled in a two-stage $^3$He system with a base temperature of 210 mK. A small magnetic field (~ 1 mT) was applied to the junction in order to suppress the Josephson tunneling current. The junctions were DC biased in the subgap region ($V_{DC}$ = 20-90 μV). The biasing circuit was a DC voltage bias, which proved to be much more stable than DC current bias, used in earlier experiments. An I-V curve with the Josephson current suppressed is shown later in Fig. 9, below. The subgap current in the region 30 μV to 80 μV is about 25 nA, which is the BCS prediction for 210 mK.[10] Fiske steps[11] are evident at $V_{DC}$ = 115 μV and 145 μV. The ability to trace out these features without hysteresis is evidence of the steep load line ($R_{load}$ ~ 10 ohms) provided by the biasing circuit. A detailed analysis of the DC and AC properties of the biasing circuit has been published.[12]

The samples were illuminated with an $^{55}$Fe x-ray source, with a dominant energy emission at 5.89 keV. Experiments have also been performed in our group with visible photons, toward the purpose of developing imaging optical spectrometers; these experiments are discussed elsewhere.[13] X-ray photons create a larger tunneling current, however, and thus give a large signal for studying quasiparticle dynamics. The effects of a heated energy distribution are also more evident with larger energy photons. The x-rays are absorbed in the Ta film and break Cooper pairs to create excess quasiparticles. The excess quasiparticles diffuse throughout the Ta absorber, are trapped in the Al electrodes, and tunnel through the oxide barrier, where they cause a temporary increase in the current. An example of such a current increase or "current pulse" is shown later in the paper, in Fig. 7. The device is uniformly illuminated, so pulses are obtained from all absorption locations with roughly equal probability in a given data set. Typically we take several thousand pulses in one data set.

The current pulses are amplified by a low noise current amplifier, digitized by an oscilloscope and stored on disk. The low noise current amplifier is formed by a 2SK147 JFET, an Amptek A250 transresistance amplifier, and additional circuitry that allows the amplifier to be DC coupled.[12] The amplifier obtains a voltage noise of 0.5 nV/sqrt (Hz), a current noise of 0.2 pA/sqrt (Hz), and a 3 dB bandwidth of 50 kHz.[12] The pulses are initially recorded without filtering, which allows us to extract the physical information from the shape of the pulse without distortion. Filtering can also be performed numerically on the saved waveforms if one wants to measure the energy resolution. One can also integrate each current pulse numerically to obtain the charge.

Each photon causes *two* current pulses, one in each junction. The sum of the charges from the two pulses, $Q_1+Q_2$, is proportional to the total number of quasiparticles created by the photon. This sum is also proportional to the photon energy, allowing the detector to perform as a spectrometer. The ratio of the two charges, $Q_1/Q_2$, can be used to extract the location of the absorbed photon, giving the device inherent imaging capability. This ability to know the absorption location of each photon is what allows the detector dynamics to be studied in such detail. Typically one plots $Q_1$ versus $Q_2$, an example of which is shown in Fig. 5. In addition to the area of each pulse there is information in the shape of each pulse: the peak current, the rise time and the fall time, and timing information between the two pulses. We measure these pulse parameters as a function of the device operation conditions (temperature, DC bias voltage, impedance environment) in order to study the device physics.

--

## III. Theory and Modeling

In this section we first list all the important physical processes that determine the current pulse and then derive a numerical model to be compared to the measured data. Fig. 2 shows a band diagram of the device and labels the important physical processes. Here we use the excitation representation, where the horizontal axis is the location (in the $x$-direction) of the quasiparticle and the vertical axis is its energy. The larger gap Ta absorber ($\Delta_{Ta}$ = 700 µV) is shown in good contact with the smaller gap Al tunnel junctions ($\Delta_{Al}$ = 170 µV). The tunnel barrier is also indicated.

The first important process is quasiparticle creation, whereby the incident photon's energy is converted into excess quasiparticles that cool to energies near $\Delta_{Ta}$ in a timescale of order several ns. This process has been described and modeled in other work.[14] In Ta it is found that approximately 60% of the energy is converted into excess quasiparticles while 40% goes into subgap phonons whose energy is insufficient to further break any Cooper pairs. These are lost into the substrate. The second important process is quasiparticle diffusion, where the quasiparticles diffuse laterally in the absorber to either tunnel junction. When the quasiparticles reach the trap region, of lower energy gap, they can scatter inelastically, emitting a phonon. The inelastic scattering causes quasiparticle trapping, whereby the quasiparticles are then confined in the Al region near the tunnel barrier, and quasiparticle thermalization, where the quasiparticle energy distribution slowly equilibrates to the lattice temperature in Al. Quasiparticle multiplication can occur if the emitted phonons in the trap break additional Cooper pairs in the Al trap.

Once inside the Al trap the process of quasiparticle tunneling is important and results in the current signal detected by the amplifier. Quasiparticles can tunnel as either electrons or holes, transferring a negative or positive charge, respectively.[15] Once a quasiparticle tunnels, it can then tunnel from the counter-electrode back to the trap, also as an electron or hole. Quasiparticle recombination, where two excess quasiparticles combine to form a Cooper pair and emit a phonon, removes quasiparticles from the tunneling region. Recombination or other losses in the absorber can also occur.[5] Quasiparticle recombination in the Al can occur between two excess, photon-induced quasiparticles ("self-recombination") or between an excess quasiparticle and a thermally

--

excited quasiparticle ("thermal recombination"). Quasiparticle outdiffusion is where a quasiparticle diffuses from the counter-electrode into the wiring leads, thus removing it from the tunneling region. If recombination phonons, of energy $E_{phonon} > 2\Delta_{Al}$, do not leave the junction area they can break Cooper pairs and reform two quasiparticles, in a process known as phonon trapping.

To fully determine the size and shape of the current pulse, one would in general want to know the time evolution of the spatial location *and* energy of each excited quasiparticle in the device. This would lead to a very complicated analysis, so we have made two important simplifications in our model: we ignore the energy distribution in the absorber and ignore spatial effects in the junction electrodes. Both are fairly good approximations. The inelastic scattering times in the tantalum absorber are relatively fast,[16] so effects from a non-zero quasiparticle energy spread in the absorber should be small. The trap has a much smaller volume than the absorber and aluminum has a much larger diffusion constant than tantalum, so spatial effects in the junction should also be minimal. The lateral diffusion in the absorber and the energy dependence of the quasiparticles in the trap are the major effects in determining the signal, for our materials and geometry.

In Fig. 3 we draw a schematic of the current pulse calculation. The calculation consists of two parts, one in the absorber and one in the junction. In the absorber we calculate the spatial distribution of quasiparticles as a function of time using the diffusion equation. The current that flows out of the absorber is calculated through a boundary condition that allows us to include the effects of quasiparticle trapping. This current, called the interface current ($I_{int}$), must flow into the trap electrode of the junction, where it is the input to the second part of the calculation. In this part we calculate the time evolution of the quasiparticle energy distribution. The total tunneling current can then be computed from this distribution. We describe the equations for each of these steps below.

Diffusion in the absorber is modeled by the one-dimensional (1-D) diffusion equation. We reduce the three dimensional diffusion in the absorber to a single dimension, the x-direction in Fig. 1. This is valid assuming there are negligible losses at the surfaces and edges in the other two directions.[7] Surface or edge loss would show up

--

experimentally as a larger energy width for photons absorbed at the center of the Ta absorber, or as an increase in the loss rate in the Ta absorber, ($\tau_{loss}^{-1}$; see below). Neither of these is evident in our experiments.[4,5,6] We are therefore confident that a 1-D treatment is appropriate. We use the 1-D diffusion equation with loss to describe the spatial and temporal evolution of the quasiparticle density $U(x,t)$:

$$\frac{\partial U}{\partial t} - D_{Ta}\frac{\partial^2 U}{\partial x^2} + \frac{U}{\tau_{loss}} = 0. \tag{1}$$

Here $D_{Ta}$ and $1/\tau_{loss}$ are the diffusion constant and loss rate, respectively, for quasiparticles in the Ta film. We treat equation (1) numerically with the Crank-Nicholson formalism.[17] The density $U$ is approximated on a spatial grid of increment $dx$; the density at a given time $t$ is represented by a column vector $\overline{U}(t)$. The density at time $t+dt$ is related to the density at time $t$ by:

$$\overline{\overline{A}}\,\overline{U}(t+dt) = \overline{\overline{B}}\,\overline{U}(t), \tag{2}$$

where $\overline{\overline{A}}$ and $\overline{\overline{B}}$ are tri-diagonal matrices. We define $\lambda = Ddt/(dx)^2$; then the matrix $\overline{\overline{A}}$ has $(1+\lambda+1/\tau_{loss})$ as its diagonal element and $-\lambda/2$ as its off-diagonal elements, while $\overline{\overline{B}}$ has $(1+\lambda-1/\tau_{loss})$ as its diagonal element and $\lambda/2$ as its off-diagonal elements. This method is unconditionally stable and is accurate to first order in both space and time.[17] A Gaussian spatial distribution with area equal to the initial charge created is the initial condition, as shown in Fig. 3.

We do not treat explicitly the energy distribution of the quasiparticles in the Ta absorber produced by photon absorption. These very rapidly cool by electron excitation and phonon emission to near the energy gap of Ta, and then diffuse at this energy. This process is fast compared to other timescales because the electron phonon coupling in Ta is strong. The time for this process is of order 0.1 ns, during which the quasiparticles spread about 1 µm in either direction. This results in a "hotspot volume" of about 2 µm x 2 µm x 0.6 µm, where the 0.6 µm is the thickness of the film. The number of excess quasiparticles over this volume is not enough to significantly depress the energy gap in Ta.[18] The quasiparticles are produced in this small volume and reach nearly the Ta gap energy before significant diffusion occurs. This situation differs from that described in

--

Ref. 8,[8] where a sandwich structure of Ta/Al films forms the absorber and tunnel junction. For this structure, the fast equilibration occurs in the volume from which the tunneling current originates. Thus, for that situation a treatment of the non-equilibrium quasiparticles in the absorber is particularly important. It is of less importance for our devices where equilibration to near the Ta gap occurs prior to trapping and tunneling.

Quasiparticle trapping occurs when quasiparticles inelastically scatter in the Al trap to an energy that is below the gap of Ta. They are then confined to the Al trap region. The number of quasiparticles created by a photon is not enough to depress the energy gap in Al.[19] We model the trapping with a boundary condition at the absorber-trap interface. We assume that the diffusion current is continuous across the interface. Therefore we can write:

$$D_{Ta} \frac{\partial U}{\partial x}\bigg|_{int,Ta} = D_{Al} \frac{\partial U}{\partial x}\bigg|_{int,Al}, \qquad (3)$$

where $D_{Al}$ is the diffusion constant in Al and the derivatives are evaluated on the Ta and Al side of the interface, respectively. In the Al trap the quantity $U$ represents the distribution of quasiparticles *above the gap of Ta*. Quasiparticles scatter below this energy in a time given by $\tau_{trap}$. Assuming the trap is semi-infinite in extent [16], the spatial distribution of quasiparticles above the gap of Ta in the Al trap then decays exponentially with a diffusion decay length given by $l_{trap} = \sqrt{D_{Al}\tau_{trap}}$. This allows us to write:

$$\frac{\partial U}{\partial x}\bigg|_{int,Al} = \frac{U(x=int,Al)}{l_{trap}}. \qquad (4)$$

Comparing (3) and (4) allows us to solve for the spatial derivative of $U$ at the Ta interface, which we write as:

$$\frac{\partial U}{\partial x}\bigg|_{int,Ta} = \frac{U(x=int,Al)}{(D_{Ta}/D_{Al})l_{trap}} = \frac{U(x=int,Ta)}{l^*_{trap}}, \qquad (5)$$

with $l^*_{trap}$ the effective trapping length. We have also assumed the density $U$ is continuous across the interface, i.e. $U(x=int,Al) = U(x=int,Ta)$. This continuity holds since we are only considering non-equilibrium quasiparticles above the gap of Ta. Equation (5) is valid for a trap which is semi-infinite in length. In our devices $l^*_{trap}$ ~ 10

--

µm. This is much less than the size of the trap, which is of order 50 µm, considering the quasiparticles have to diffuse the length of the junction in order to exit the trap (see Fig. 1). This *conservatively* assumes that the trapping occurs *only* in the bulk Al and not in the "overlap" region between the Ta absorber and the Al trap. We show below that this is indeed the case.

To implement (5) into the model, the derivative on the left hand side of (5) is approximated numerically by taking its finite-difference. Then it can be substituted into the last two columns of the matrices $\bar{\bar{A}}$ and $\bar{\bar{B}}$ in equation (2). The quasiparticle current that is trapped, $I_{int}$, is given by either side of equation (3).

The current $I_{int}$ flows into the Al trap at an energy of $\Delta_{Ta}$. Once inside the junction the quasiparticles can scatter inelastically to lower energy, tunnel to the counter-electrode, or recombine to form Cooper pairs. In the counter-electrode quasiparticles can inelastically scatter, tunnel back to the trap, out-diffuse into the wiring leads, or recombine. The relative timescales for these processes determine the time evolution of the quasiparticle energy distribution on each side of the barrier and ultimately the tunneling current. We solve for this time-dependent energy distribution through a system of rate equations. The trap and counter-electrode are divided up into energy intervals of size $2\delta$, indicated in Fig. 3. We define the quantities $N_{tr}[E_i]$ and $N_{ce}[E_i]$ as the number of quasiparticles in the energy interval $(E_i+\delta) - (E_i-\delta)$ in the trap and counterelectrode, respectively. Below we detail the time evolution of $N_{tr}[E_i]$ and $N_{ce}[E_i]$. The equations for $N_{tr}[E_i]$ and $N_{ce}[E_i]$ are sometimes the same, so in such cases we write down only one equation using the variable $N[E_i]$, understanding that it corresponds to *two* equations, one where $N[E_i]$ represents $N_{tr}[E_i]$ and one where $N[E_i]$ represents $N_{ce}[E_i]$. The rates for the various processes will be different numbers for the trap and the counterelectrode and depend on geometry.

In general the time rate of change of $N[E_i]$ is given by:

$$\frac{dN[E_i]}{dt} = \frac{\partial N[E_i]}{\partial t}\bigg|_{I_{int}} + \frac{\partial N[E_i]}{\partial t}\bigg|_{scat} + \frac{\partial N[E_i]}{\partial t}\bigg|_{tun} + \frac{\partial N[E_i]}{\partial t}\bigg|_{rec} + \frac{\partial N[E_i]}{\partial t}\bigg|_{out}. \quad (6)$$

The terms on the right hand side represent, in order, the change in *N* due to the interface current (trap only), scattering, tunneling, recombination, and outdiffusion

--

(counterelectrode only). Here we have only included terms which give first-order effects in the current pulse. This is done in order to keep the number of adjustable parameters in the model equal to the number of measurements in our experiments. *Thus, all parameters can be determined from experiment.* Higher order processes such as phonon absorption by a quasiparticle, quasiparticle recombination with energy exchange, and effects due to local quasiparticle traps have been ignored because they give minimal impact to the output tunneling current. To implement (6) we divide up the total energy range between $\Delta_{Ta}$ and $\Delta_{Al}$ into $M$ intervals; then we solve the $2M$ coupled differential equations to determine the time evolution of the system. We describe each term on the right hand side of equation (6) separately.

The first term on the right hand side of (6) is the change due to quasiparticles entering from the absorber. Quasiparticles enter only on the trap side at energy $\Delta_{Ta}$. Thus the only non-zero term comes from the trap side of the junction at energy $E_M$:

$$\left.\frac{\partial N_{tr}[E_M]}{\partial t}\right|_{I_{int}} = \frac{I_{int}}{e}. \tag{6a}$$

Here $e$, the value of the electron charge, is defined as a positive number. The second term in (6), the change in $N$ due to inelastic scattering, is given by the number that scatter into energy interval $E_i$ from higher energies minus the number that scatter from energy $E_i$ to lower energies. This term is the same for both the trap and the counterelectrode:

$$\left.\frac{\partial N[E_i]}{\partial t}\right|_{scat} = \sum_{j=i+1}^{M} \frac{N[E_j]}{\tau_s[E_j,E_i]} - \sum_{k=0}^{i-1} \frac{N[E_i]}{\tau_s[E_i,E_k]}. \tag{6b}$$

Here $1/\tau_s[E_a,E_b]$ is the rate to scatter from a given energy $E_a$ to energy $E_b$. These rates are computed using the expressions in ref. 15 and depend only on the energies and a material dependent prefactor $\tau_0$, to be discussed later. The third term is due to tunneling, and differs on the trap and counterelectrode side. On the trap side we have:

$$\left.\frac{\partial N_{tr}[E_i]}{\partial t}\right|_{tun} = -\frac{N_{tr}[E_i]}{\tau_{tun}^{tr}[E_i+eV]} - \frac{N_{tr}[E_i]}{\tau_{tun}^{tr}[E_i-eV]} + \frac{N_{ce}[E_i+eV]}{\tau_{tun}^{ce}[E_i]} + \frac{N_{ce}[E_i-eV]}{\tau_{tun}^{ce}[E_i]}. \tag{6c}$$

Here $1/\tau_{tun}[E]$ is the tunneling rate *into* energy $E$, with the superscripts "tr" and "ce" indicating which side the quasiparticle is tunneling *from*; $V$ is the DC voltage across the junction and $e$ is the electron charge. The terms in (6c) represent, in order, electron

-12-

--

tunneling from the trap, hole tunneling from the trap, electron tunneling from the counter-electrode, and hole tunneling from the counter-electrode. Electron tunneling raises the quasiparticle energy by $eV$ in going across the barrier; hole tunneling lowers it by $eV$. Besides the final energy, the tunneling rate depends on the density of states, the electrode volume, and the opacity of the barrier. Expressions for the tunneling rates and their energy dependence can be found in the literature.[20] For the counter-electrode we have:

$$\left.\frac{\partial N_{ce}[E_i]}{\partial t}\right|_{tun} = -\frac{N_{ce}[E_i]}{\tau_{tun}^{ce}[E_i+eV]} - \frac{N_{ce}[E_i]}{\tau_{tun}^{ce}[E_i-eV]} + \frac{N_{tr}[E_i+eV]}{\tau_{tun}^{tr}[E_i]} + \frac{N_{tr}[E_i-eV]}{\tau_{tun}^{tr}[E_i]}. \quad (6d)$$

The fourth term is due to recombination, and is the same on both sides:

$$\left.\frac{\partial N[E_i]}{\partial t}\right|_{rec} = -\frac{R^*}{V}\left(\sum_{j=1}^{M} N[E_j] + 2n_{th}V\right)N[E_i]. \quad (6e)$$

Here $V$ is the volume of the electrode in question, $R^*$ is the recombination rate per unit density of quasiparticles and $n_{th}$ is the thermal density of quasiparticles, which is a temperature-dependent quantity. The recombination rate is given by $R^*$ instead of $R$ to account for an enhancement by phonon-trapping, in the usual fashion.[20] In theory $R^*$ should be an energy-dependent quantity and be replaced by a matrix $R^*_{ij}$, similar to the scattering matrix in (6b). In practice we find that replacing the energy dependence with a single, average $R^*$ is a very good approximation; this is both because the energy dependence is somewhat weak and because the overall recombination is not so strong at the temperatures and time scales of interest. The first term on the right hand side of (6e) is the self-recombination term and the second is the thermal recombination term. Expressions for $n_{th}$ and $R^*$ can be found in ref. 20;[20] they depend on the energy gap and critical temperature in the Al electrode, the bath temperature, the density of states and $\tau_0$.

The final term on the right hand side of (6) is the loss of quasiparticles due to outdiffusion from the counterelectrode:

$$\left.\frac{\partial N_{ce}[E_i]}{\partial t}\right|_{out} = \frac{N_{ce}[E_i]}{\tau_{out}}, \quad (6f)$$

where $1/\tau_{out}$ is the rate of outdiffusion.

The system of $2M$ differential equations (6) is solved by the modified Euler method[17] because only full time steps are used; Runga-Kutta methods were not chosen

-13---

because it required evaluation at half-time steps, and performing the absorber part of the calculation at half-time steps was costly in run time. Having solved for the full distributions $N_{tr}[E_i]$ and $N_{ce}[E_i]$ as a function of time the current can be calculated, which is just the number $N_{tr}[E_i]$ or $N_{ce}[E_i]$ times the tunneling rate for each interval. Hole tunneling from the trap and electron tunneling from the counterelectrode contribute a positive current while hole tunneling from the counter-electrode and electron tunneling from the trap contribute a negative current. Defining current as the flow of positive charge we can then write:

$$\frac{I_{tun}}{e} = \sum_i \left( \frac{N_{tr}[E_i]}{\tau_{tun}^{tr}[E_i - eV]} - \frac{N_{tr}[E_i]}{\tau_{tun}^{tr}[E_i + eV]} + \frac{N_{ce}[E_i]}{\tau_{tun}^{ce}[E_i - eV]} - \frac{N_{ce}[E_i]}{\tau_{tun}^{ce}[E_i + eV]} \right). \quad (7)$$

Here the terms represent contributions to the tunneling current from, in order: hole tunneling from the trap, electron tunneling from the trap, hole tunneling from the counter-electrode, electron tunneling from the counter-electrode. The time evolution of (7) is the output of the model and can be compared to the experimental data.

## IV. Quasiparticle Dynamics

In this section we compare predictions of the model to experiment and discuss the values of the fitting parameters. We need three types of inputs to the model: (1) experimental constants, (2) physical constants, and (3) fitting parameters. The experimental constants include the absorber length, the trap and counter-electrode volume, the operating temperature, the junction DC voltage, the junction normal resistance and the energy gap in Al. They are fixed by the device geometry and experimental conditions or, in the case of the junction resistance and energy gap, extracted from DC measurements of the I-V curve. The physical constants include the density of states at the Fermi Surface in Al, $D(E_f)$, the critical temperature of Al, and Kaplan's electron-phonon time in Al ($\tau_0$).[16] They are extracted from other measurements in the literature,[21] although we will allow the value of $\tau_0$ to vary somewhat. The fitting parameters are the diffusion constant in Ta ($D_{Ta}$), the trapping time ($\tau_{trap}$), the absorber loss time ($\tau_{loss}$), the trapped charge ($Q_0$) and the outdiffusion time ($\tau_{out}$). We emphasize that all five fitting parameters are constrained by *independent* measurements of the

-14-

--

current pulses. We will discuss these five parameters and their associated measurements and then fit the time-dependent pulses. We recall that for each photon we can extract its absorption location along the absorber through the ratio of the two charges.

| Fitting Parameter | Measurement type | Theory | Experiment |
|---|---|---|---|
| Diffusion Constant ($D_{Ta}$) | Delay time | 40 cm$^2$/s | 8 cm$^2$/s |
| Trapping time ($\tau_{trap}$) | Charge division | 6 ns | < 10 ns |
| Absorber loss time ($\tau_{loss}$) | Curvature | 2.8 ms | 31 µs (89 µs in [5]) |
| Trapped charge ($Q_0$) | Peak current | 8x10$^6$ e$^-$ | 8x10$^6$ e$^-$ |
| Outdiffusion time ($\tau_{out}$) | Total charge | 5-10 µs | 7.1 µs |

Table 1: Fitting parameters, measurement type and results

The five fitting parameters, their associated measurements, and theoretical and experimental values are shown in table 1. The first parameter, the Ta diffusion constant, is determined from the difference in the arrival times of the current pulses in the two junctions, at a specific current threshold. This measurement is shown in Fig. 4, where we plot this delay time versus absorption location (two current pulses from a single photon are shown in Fig. 7, where one can easily see the delay in arrival times). The data are compared with results from the simulation. A good fit for two different threshold currents is obtained for $D_{Ta} = 8$ cm$^2$/s. This value is much lower than the value we calculate using the low temperature resistivity of our Ta, 0.48 µΩ-cm, which would predict $D_{Ta} = 40$ cm$^2$/s. If we account for a slowdown due to a reduction in quasiparticle group velocity,[22] this value still only reduces to 27 cm$^2$/s. The low experimental value has been discussed previously and similar results have been found in other work, including materials other than Ta.[5,23] The physical origin of the slow diffusion of quasiparticles remains an open question in the tunnel junction detector field.

In Fig. 5 we plot the two charges $Q_1$ and $Q_2$, versus each other, for two different operating temperatures. Such a plot can reveal much of the device physics, as has been shown in other work.[7] With no loss the plot would be a straight line for fixed photon energy; some losses are evident in the graph, discussed below. Higher energies appear as displaced lines, such that ($Q_1 + Q_2$) is larger. One can see the stronger $\alpha$ line (5.89 keV) and the weaker $\beta$ line (6.49 keV) for each temperature. Near the edges the average

--

charge is higher due to absorptions in the Al trap, where the energy gap is lower. The fits from the model, obtained by generating current pulses for each location and integrating the charge, are also shown, with good agreement.

The strength of the trapping can be inferred from the $Q_1$ versus $Q_2$ by focusing on events close to one junction, where the charge in one junction is large and the other is small. If the trapping is fast (small $\tau_{trap}$) quasiparticles are trapped immediately and the charge in the opposite junction is nearly zero. This will cause the points to extend to both axes. If the trapping is slow (large $\tau_{trap}$), the points will cluster toward the center of the graph; this is because for events near the edges, quasiparticles that are not trapped can diffuse back into the absorber and cause a finite charge in the other junction. The division of charge in our data can be fit with a trapping time of $\tau_{trap} < 10$ ns. The predicted scattering time for a quasiparticle in Al at the energy of Ta is 6 ns.[16] The measurement is not sensitive enough to confirm a more precise agreement. The fact that $\tau_{trap}$ is not significantly longer than 6 ns suggests very little obstruction of transport at the Ta/Al interface, indicating that the cleaning of the Ta interface prior to deposition is effective.

The value of 10 ns also strongly suggests that the quasiparticle trapping occurs in the bulk Al, and not in the Ta/Al overlap region. In the overlap region the average value of the energy gap is higher than in the Al itself. Thus, the inelastic (electron phonon) scattering time would be much *larger* in the overlap region, of order hundreds of ns. This would be incompatible with the 10 ns trapping time we observe. In addition, if fast trapping occurred inside the overlap region, the $Q_1$ vs. $Q_2$ plot would be different. Events absorbed at the end of the Ta absorber, in this overlap region, would not be able to cause even a small signal in the other junction. This would cause the main sequence of events to extend all the way to the axes, which is not observed. These facts help confirm the validity of using the stated boundary condition in equation (5).

The amount of loss in the absorber can be inferred from the degree of curvature in the $Q_1$ versus $Q_2$ plot. Quasiparticles created in the center of the device have to diffuse a longer average distance/time, and are more susceptible to loss. Since the diffusion constant is constrained by the delay time measurement, we vary the value of $\tau_{loss}$ to fit the data. The value we find for these devices is $\tau_{loss} = 31$ μs. Experiments on newly fabricated devices find a larger value of $\tau_{loss}$ = 82 μs; the difference between the two values is believed to be due to a modification of the fabrication process, discussed in ref. 5. Both of these values fall well below of the theoretical value for the loss due to thermal recombination in Ta, 2.8 ms.[16] We speculate that the loss is due to local depressions in the energy gap, causing quasiparticles to be confined until they eventually recombine.[23] Such depressions can result at the surfaces or due to perpendicular magnetic flux penetration. The latter can occur if there is a small misalignment in the parallel field required to suppress the Josephson current. These losses limit the size of the absorber one can eventually use.[5]

With the values of $D_{Ta}$, $\tau_{loss}$ and $\tau_{trap}$ fixed, the shape of the $Q_1$ versus $Q_2$ plot (for a given energy) is determined. However, the magnitude of the charge that tunnels, $Q_1$ + $Q_2$, depends on two other parameters. The first is the trapped charge, $Q_0$, which is the charge that enters the junction. Note that this is not necessarily equal to the charge created by the photon (see below). The second is the outdiffusion time, $\tau_{out}$. Quasiparticles that do not diffuse away can tunnel multiple times, adding to the total charge. We can distinguish between these two by the shape of the pulse (see Fig. 7). Charge that initially enters the junction ($Q_0$) tunnels early in the current pulse, and thus affects the value of the peak current. Charge that continues to tunnel due to slow outdiffusion ($\tau_{out}$) appears near the end of the pulse and does not affect the peak current. Thus we vary $Q_0$ to fit the values of the peak current and then vary $\tau_{out}$ to fit the total, tunneled charge ($Q_1$ + $Q_2$).

In Fig. 6 we plot the peak current in one junction, $I_{P1}$, versus the peak current in the other junction, $I_{P2}$, for a data set. A best fit from the model is shown with a value of $Q_0$ = 8x10$^6$ electrons. The value $Q_0$ is a product of the initially created charge, $Q_{cr}$, and the multiplication-upon-trapping factor, γ. Our simulations of the trapping find that γ = 1.6 +/- 0.2. The value of $Q_{cr}$ is estimated by Zhender, who predicts 40% of the energy should go into phonons; this gives $Q_{cr}$ = 0.6($E_{x-ray}/\Delta_{Ta}$) = 5x10$^6$ electrons.[14] Combining these two we predict a value of $Q_0$ of 8x10$^6$ electrons, in good agreement. However, the experiment is only sensitive to the total trapped charge; it cannot yet differentiate between the mechanisms of quasiparticle creation and multiplication upon trapping.

--

The remaining fit parameter is the outdiffusion time, which we adjust in the model to fit the total charge ($Q_1 + Q_2$) after the value of $Q_0$ has been set. The fits are shown in Fig. 5. We find a good fit for an outdiffusion time of 7.1 μs. The complex shape of the junction and the wiring (Fig.1) makes $\tau_{out}$ somewhat difficult to estimate, since we do not know exactly what average length the quasiparticles must diffuse before they can no longer tunnel. A rough estimate would be 200 μm from the center of the junction, which would make $\tau_{out} = (200\ \mu m)^2/D_{Al} = 6.7$ μs, in reasonable agreement with the data. Note that the diffusion constant in Al ($D_{Al} = 60$ cm$^2$/s, measured in a separate device[4]), like $D_{Ta}$, is also smaller than expected.

We have used events from the entire absorber to give a best-fit value for the fitting parameters; what remains to be done is to look at the time dependence of a *single* pair of current pulses to see if the time-dependent current from the model fits the observed waveforms. This is shown in Fig. 7, where we show the two pulses from a single absorption event and their fits. All the inputs to the model at this point are constrained; there are no adjustable parameters at this point. The excellent agreement obtained in Fig. 7 confirms our diffusion-tunneling model and gives us confidence for further exploration of the device physics.

**V. Quasiparticle Energy Distribution**

In this section we explore the effects of the quasiparticle energy distribution in the junction electrodes. When the quasiparticles enter the Al trap they do so at an energy of $\Delta_{Ta} = 700$ μV $= 4.1\Delta_{Al}$. The time for a quasiparticle to scatter below this energy, emitting a phonon, is the trapping time. As we saw above, $\tau_{trap} < 10$ ns both in theory and in our measurements. This is much faster than the tunneling time, about 2.4 μs in our device. One might conclude from this that the quasiparticles will completely thermalize before they tunnel, having all scattered to an energy within approximately times $kT_{bath}$ of the Al gap. However, $\tau_{trap}$ is the time for a quasiparticle to scatter to *any* energy below $4.1\Delta_{Al}$, not necessary to within an energy of $kT_{bath}$ above $\Delta_{Al}$. The time for the whole distribution to thermalize is much longer. This is because as quasiparticles reach lower energies the scattering time is longer, due to the decreasing phase space for phonon emission. Hence quasiparticles which do not scatter to within $kT_{bath}$ of $\Delta_{Al}$ in their first scattering event will

-18-

--

survive much longer at their new energy and the full thermalization will take longer than the initial 10 ns.

In order to gain insight into the thermalization process we have done a computer simulation of the time dependence of the energy distribution. We look at a subset of the full current model to focus just on the inelastic scattering in the trap electrode. We start with $5 \times 10^6$ quasiparticles at $4.1 \Delta_{Al}$ in the Al trap and use the set of equations in (6b) to solve for the time evolution of the energy distribution. We ignore the counterelectrode entirely and the other terms in (6). The results are shown in Fig. 8, where we show the quasiparticle concentration as a function of energy, at 0.1 µs and 1 µs. From the graph we can see there are still a significant number of quasiparticles in the range 70-100 µeV even after 1 µs. Given that $kT_{bath} = 17$ µeV in our experiments, the assumption of a thermal distribution while tunneling is incorrect. We have included no phonon absorption by quasiparticles in the calculation, which would only increase the average quasiparticle energy. The essential point of Fig. 8 is that given the rates for phonon emission, there is simply not enough time for the distribution to thermalize before the quasiparticles start tunneling.

This energy distribution for quasiparticles in the trap is inherently non-equilibrium and does not have the shape of a thermal distribution. In our model we *explicitly* keep track of the time-evolution of this non-equilibrium distribution. Nevertheless for the purposes of discussing effects related to the circuit impedance and noise, it is often useful to represent the distribution with a single number for the time of peak current flow, the tunnel time. In that case we characterize the distribution with an effective temperature, $T_{eff}$, where $T_{eff} > T_{bath}$. The value of $T_{eff}$ is chosen by finding the temperature whose thermal I-V curve (normalized to the number of photon-induced quasiparticles) has the same slope as the I-V curve at the bias point during the peak of the tunneling pulse (see below). We estimate $T_{eff} = 0.7\text{-}0.8$ K.

The high value of $T_{eff}$ has a major effect on the voltage dependence of the tunneling current. In Fig. 9 we show the I-V curve for two conditions: (i) the quiescent state, with no excess tunneling current; and (ii) the dynamic state, shortly after photon absorption, where there are extra quasiparticles on the trap side at a temperature of $T_{eff} = 0.7$K. The first curve is the actual DC current measured by our electronics; the second

--

curve is a calculation of the expected behavior. We observe that a typical bias voltage of 70-80 µV is in the flat region of the I-V curve in the quiescent state (low conductance/high resistance), but moves to a region of larger slope (higher conductance/lower resistance) during the photon pulse. This increase in slope is larger than the expected increase due solely to the extra current flowing. The relationship between the quasiparticle energy distribution and the I-V curve has been discussed in detail previously.[15] The fact that there are quasiparticles at energies higher than a few times $kT_{bath}$ allows for more transfer of charge as holes, against the bias, and this causes the increase in slope.[4,15] The increase in slope has two effects: the junction conductance is increased during tunneling, and the total collected charge is now a stronger function of bias voltage. We now demonstrate these effects experimentally.

We can first estimate the conductance of the junction during the pulse using the model developed in section III. We assume a DC bias voltage of 80 µV, where the quiescent resistance, $(dI/dV)^{-1}$, is approximately 15 kΩ. We run the tunneling current calculation for two different bias voltages, one at 80.5 µV and one at 79.5 µV. We consider an absorption event at the center of the absorber. At each time step we add the excess tunneling current from the calculation to the quiescent current and estimate the differential resistance from the current at these two voltages:

$$R_{eff} \approx \frac{1\mu V}{I(V_{DC} = 80.5 \mu V) - I(V_{DC} = 79.5 \mu V)} \quad . \tag{8}$$

Here I is the total current (quiescent plus excess). The results are shown in Fig. 10, where we plot $R_{eff}$ versus time. We can see that before the pulse the resistance is 15 kΩ, but drops to less than 3 kΩ at the peak of the pulse. The observed peak current for an event from the center is about 55 nA; the quiescent current is about 25 nA. If there were no change in the energy distribution, one would expect a junction resistance of 15 kΩ*(25nA/55nA) = 6.8 kΩ at the peak of the pulse. The model, which keeps full track of the non-equilibrium energy distribution, predicts a junction resistance smaller than that.

An experimental estimate of the junction resistance during a pulse can be obtained by looking at a graph of the peak current ($I_p$) versus DC bias voltage. This is shown in Fig. 11. The peak of the measured current pulse is plotted as a function of bias voltage. Two different absorption locations are shown, at the center (0 µm) and near the edge

-20-

--

(+75 µm), with fits from the model. The model once again shows good agreement. The slope of the graph, $(dI_p/dV)^{-1}$, represents the value of the junction impedance at the peak of the pulse. This value is about 2.5 kΩ for events from the center and 1.5 kΩ for events near the edge. The lower resistance near the edge is due to the larger junction current and faster collection into the trap. The 2.5 kΩ value for absorption events at the center agrees well with the minimum resistance reached by the simulation in Fig. 10 and is again much lower than expected if there were a thermal distribution of quasiparticles in the junction.

Another estimate of the junction resistance during a pulse can be made by adding series resistance to the junction, thus reducing the amount of charge collected by the amplifier. Fig. 12a illustrates the idea. The junction is modeled as a current source in parallel with $R_{eff}$. The current amplifier and any added series resistance ($R_s$) are in parallel. Physically $R_s$ corresponds to a variable resistor added in series with the amplifier. The current amplifier looks like a low impedance, around 100 Ω.[12] In Fig. 12b we plot the total collected charge versus $R_s$, and fit the charge reduction with the simple resistive division indicated by Fig. 12a. We find fitted values of $R_{eff}$ = 3.06 kΩ for events from the center and 1.55 kΩ for events from the edge, in agreement with the above measurements. We also note from Fig. 12b that the charge reduction is different for different locations in the absorber. Events from the edge suffer more reduction due to their lower $R_{eff}$. This has the opposite effect as quasiparticle losses in the absorber, which reduce the charge more for events in the center. In Fig. 13 we show plots of $Q_1$ versus $Q_2$ for series resistances of 100Ω and 400Ω. In the 400Ω case the plot appears to "straighten" out. This follows from Fig. 12b, because the events from the edge have more reduction than events from the center. The fact that simply adding series resistance can change the apparent curvature is a striking effect. Since the curvature is used as a measure of the quasiparticle loss time, one should be careful to check the series impedance of the circuit when inferring the loss time in similar devices. A similar effect on the $Q_1$ versus $Q_2$ has also been seen versus bias voltage.[24]

With a lower junction resistance the amplifier voltage noise should also matter more for measurements of the noise and energy resolution. These effects have also been discussed previously, and in fact are one of the major limiting factors in the energy resolution of these tunnel junction detectors.[6]

The second consequence of larger slope in the dynamic I-V curve (Fig. 9) is that the collected charge is a function of bias voltage. This effect has been discussed in previous work.[4,6] A plot of the total charge versus bias voltage is shown in Fig. 14, with the fit from the model. We also include fits for values of the electron-phonon scattering time, $\tau_0$, which are two times larger and two times smaller. With a longer electron-phonon time the thermalization is slower, meaning more quasiparticles remain at higher energies and less charge is collected. With a shorter value of $\tau_0$ the scattering is faster and more charge is collected. In principle this dependence can be used to fit the value of $\tau_0$, but in practice the dependence is a bit too weak. The dependence of charge on bias voltage also leads to a noise term in the energy resolution, which has also been discussed previously.[6]

The dependence of the collected charge on temperature is also a measure of the electron-phonon time in our Al films. Two quasiparticles that recombine to form a Cooper pair must also emit a phonon; thus the recombination rate is proportional to the phonon-emission rate. The temperature dependence arises because the recombination rate scales with the density of quasiparticles. This quantity, $n_{th}$, is included in the model in equation (6e). More thermal quasiparticles cause more recombination in the trap, and hence less charge tunnels from the trap. Fig. 15 shows the total charge versus temperature with fits from the model, again for larger and smaller values of $\tau_0$. It is clear here that the value of $\tau_0$ has a stronger impact than in Fig. 14, so we can try to use this data to find a fitted value of $\tau_0$.

In order to make a plot such as Fig. 15 one must be careful to differentiate between the two different roles of $\tau_0$. At low temperatures, where there are essentially no thermal quasiparticles, the thermalization of photon-induced quasiparticles prior to tunneling is the dominant process affected by $\tau_0$. Smaller values of $\tau_0$ give more efficient thermalization and a larger collected charge. At high temperatures, where there are a significant number of thermal quasiparticles, recombination is the dominant process affected by $\tau_0$. Here smaller values of $\tau_0$ give a *smaller* collected charge due to more recombination. Curves for different $\tau_0$ must thus cross over, as seen in Fig. 15. For this

--

temperature study we are interested in *only* the effects of recombination, not thermalization. If we change the value of $\tau_0$ in the model, we change *both* the amount of thermalization and the recombination. To study the recombination alone, we remove the effects of thermalization by adjusting other parameters in the model. Since thermalization effects are independent of temperature, this is easily done.

For the different values of $\tau_0$ in Fig. 15 we have made slight adjustments to $Q_0$ in order to remove the effects of thermalization on the total charge. These adjustments are within the uncertainty of the previous fitting. We now find that the value of $\tau_0$ that best fits the temperature dependence of the charge is $\tau_0 = 0.44$ μS. This is in fact the same value calculated by Kaplan et al.[16]

One might consider this a "measurement' of $\tau_0$, but this is only partially the case, as we have not accounted for the effects of phonon trapping. When two quasiparticles recombine, they emit a phonon with energy greater than $2\Delta$. If these recombination phonons do not quickly diffuse away from the junction, they can break additional Cooper pairs and re-form quasiparticles. This can increase the effective recombination time, making the value of $\tau_0$ appear larger. We do not know the exact amount of this enhancement; a naïve estimate is a factor of two. This would predict a value of $\tau_0 = 0.22$ μS, which is smaller than the value calculated by Kaplan et al., but in closer agreement to other experiments.[2,25] Future work is necessary to make our measurement more precise. We note that our previous estimates[4] of $\tau_0$ did not account for any effects of phonon trapping.

## VI. Summary

We have performed an extensive theoretical and experimental study on our Al tunneling structures to demonstrate the effects of the dynamics and energy distribution of non-equilibrium quasiparticles. A full model calculation, including the effects of diffusion, trapping, tunneling, inelastic scattering, recombination and outdiffusion, explains the measured tunneling current from photon-induced quasiparticles. The fitting procedure is done with no adjustable parameters and yields values for the diffusion

--

constant in the absorber, the loss time in the absorber, the trapping time, the trapped charge and the outdiffusion time. Some of these values disagree with theory and provide incentive to further study these devices. Measurements of the tunneling current as a function of voltage and the series resistance of the circuit demonstrate that the quasiparticle energy distribution is at a higher effective temperature than the bath during tunneling. Combining the fits from the model with the voltage and temperature dependence of the charge we study different values of the electron-phonon time, a very important parameter for the study of electrons at low temperatures. Our fitted values are within the range of those found in other work.

The value of these experiments can be seen in both the future performance of superconducting photon detectors and the basic physics of quasiparticle transport and tunneling. The unsolved problems of the slow diffusion and the cause of losses in the absorbing film suggest more research in this area, with both new physics and improved detectors a possibility. The question of quasiparticle transport in more complex geometries can be approached by the formalism we have developed. The problem of the energy relaxation gives incentive to investigate other materials, with a smaller $\tau_0$, for detector tunneling electrodes. In addition, it emphasizes the care one must take in interpreting future quasiparticle tunneling experiments.

We thank A. E. Szymkowiak, R. J. Schoelkopf, S. H. Moseley, A. Davies, R. Lathrop, M. Devoret, D. Schiminovich and B. Mazin for useful discussion and experimental assistance. This research was supported by NASA NAG 5-5255 and NASA Graduate Fellowships to KS, CW and MG.

Figure Captions:

**Fig. 1:** Schematic of the device, top and side view. The absorbing film is tantalum, the junctions are aluminum-aluminum oxide-aluminum, and electrical contact is made with a thin strip of niobium. The device sits on an oxidized silicon wafer.

**Fig. 2:** Band diagram of the device in the excitation representation. The important quasiparticle processes are labeled: (1) Quasiparticle generation (2) Quasiparticle diffusion (3) Quasiparticle trapping (4) Quasiparticle tunneling (5) Quasiparticle recombination (6) Quasiparticle outdiffusion.

**Fig. 3:** Schematic of the current pulse calculation. Top: absorber part of the calculation. Quasiparticle diffusion is simulated to calculate the interface current ($I_{int}$) that leaves the absorber and enters each of the junctions. Bottom: quasiparticles enter either of the junctions from the absorber, where they can undergo several different processes. They can scatter to lower energies, recombine to form Cooper pairs, tunnel to the counter-electrode, back-tunnel from the counter-electrode to the trap, and outdiffuse from the counter-electrode. The rates for these processes are energy-dependent, so the electrodes are divided into energy bins. Each energy bin has a rate equation, and by solving the entire system the full junction dynamics are calculated.

**Fig. 4:** Delay time versus location for threshold currents of 2 nA and 6 nA. The solid lines show fits from the model assuming a diffusion constant of 8 cm$^2$/s.

**Fig. 5:** $Q_1$ vs. $Q_2$ for $T = 0.225$ K (higher charge) and $T = 0.312$ K (lower charge). The solid lines show fits from the model.

**Fig. 6:** Peak currents $I_{p1}$ vs. $I_{p2}$ for $T = 0.225$ K. The solid lines show fits from the model.

**Fig. 7:** Two current pulses from a single x-ray absorption event, with fits from the model. $T = 0.225$ K and V = 70 µV.

**Fig. 8:** Calculated energy distribution in the Al trap during tunneling, after 0.1 µs and 1 µs. The value of $kT_{bath}$ is 17 µeV, so it is clear that the energy distribution is hotter than thermal. Although the distribution is non-equilibrium, it can be approximated with an effective temperature of about 0.7 K.

**Fig. 9:** Junction I-V curve for the quiescent state and the dynamic state 1 µs after photon absorption. The I-V in the quiescent state is measured, whereas the one for the dynamic state is a calculated estimate. A typical operating point, shown at about 80 µV, is in the flat region of the I-V in the quiescent state but moves to a region of larger slope in the dynamic state.

**Fig. 10:** Simulation of the effective junction resistance, $R_{eff}$, during a pulse for $V_{DC} = 80$ µV and T = 0.225 K. The event is calculated to be from the center of the absorber. The value of $R_{eff}$ drops below 3 kΩ during the peak of the pulse.

--

**Fig. 11:** Peak current ($I_p$) vs. bias voltage ($V_{DC}$), for events from the center ($X_0 = 0$ μm) and near the edge ($X_0 = 75$ μm) for T = 0.225 K. The solid lines show fits from the model.

**Fig. 12:** Charge reduction cause by the low value of $R_{eff}$ during a pulse. (a) Electrical equivalent circuit, showing $R_{eff}$ in parallel with the amplifier ($R_A$) and added series resistance ($R_S$). Due to the low value of $R_{eff}$, a fraction of the x-ray current does not flow through the amplifier ($R_A$). (b) Total charge vs. $R_S$. Increasing the value of $R_S$ reduces the total charge. The data is fit with a value of $R_{eff}$ of 3.06 kΩ for events in the center and 1.55 kΩ for events on the edge.

**Fig. 13:** Full $Q_1$ vs. $Q_2$ plots for (a) $R_S = 100$ Ω and (b) $R_S = 400$ Ω. The 400 Ω plot appears to straighten out due to the increased charge reduction near the edges.

**Fig. 14:** Total collected charge vs. DC voltage, with different fitted values of $\tau_0$. The charge increases with voltage well past $V_{DC} = kT_{bath}/e$. Stronger electron-phonon coupling (smaller $\tau_0$) increases the amount of thermalization prior to tunneling and results in more collected charge.

**Fig. 15:** Total collected charge vs. temperature, with different fitted values of $\tau_0$. The value of $\tau_0 = 0.44$ μS best fits the data, but does not include the effects of phonon trapping.

--

--

**Fig. 1:**

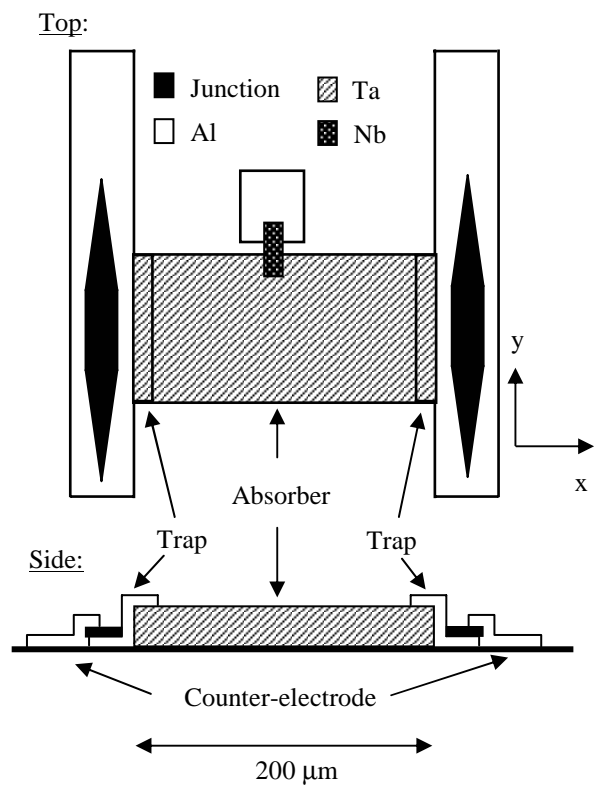

**Fig. 2:**

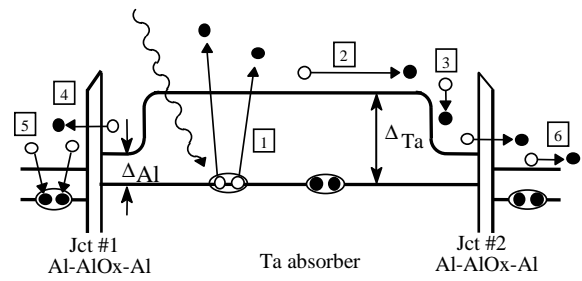

**Fig. 3:**

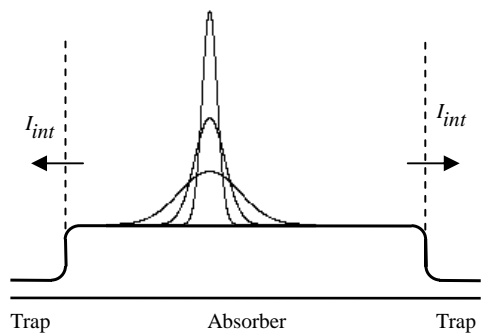

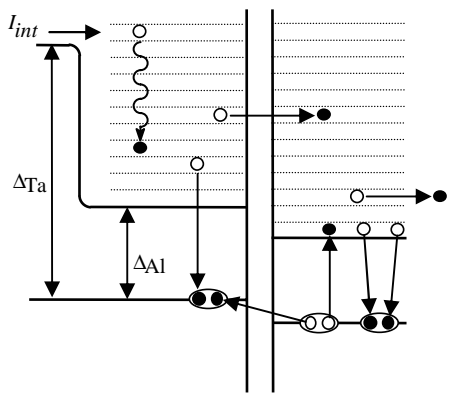

**Fig. 4:**

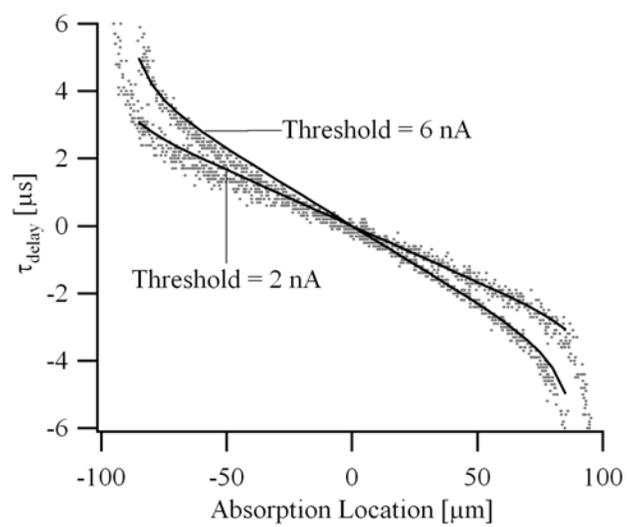

**Fig. 5:**

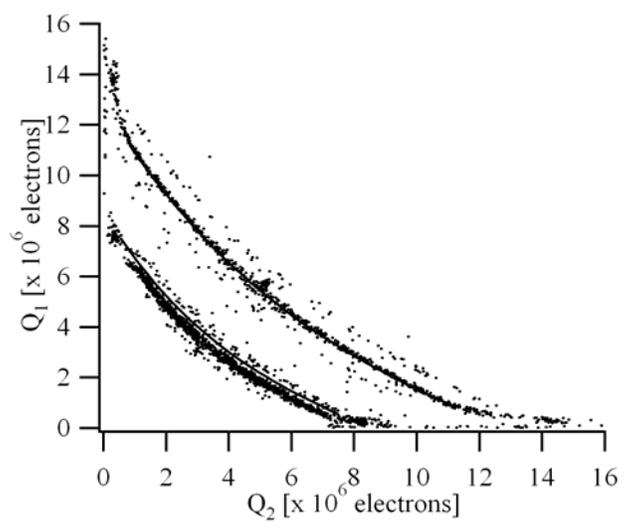

**Fig. 6:**

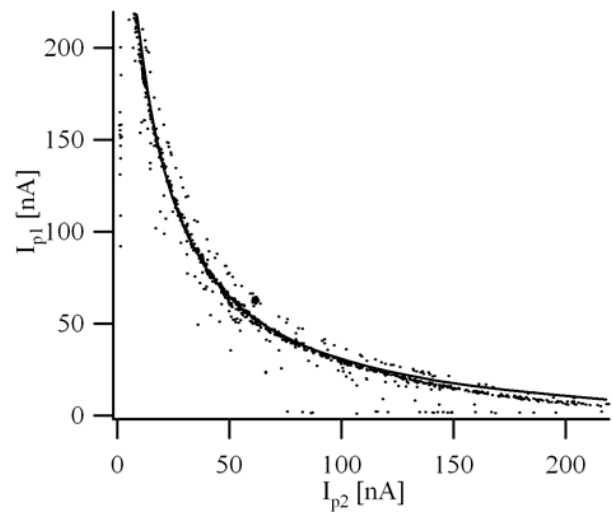

**Fig. 7:**

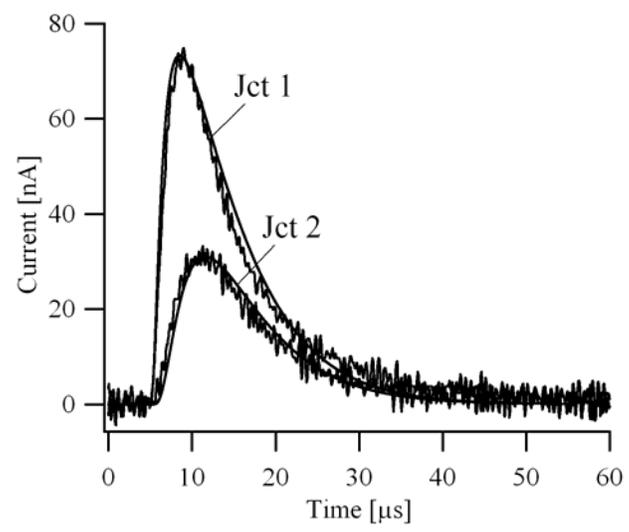

**Fig. 8:**

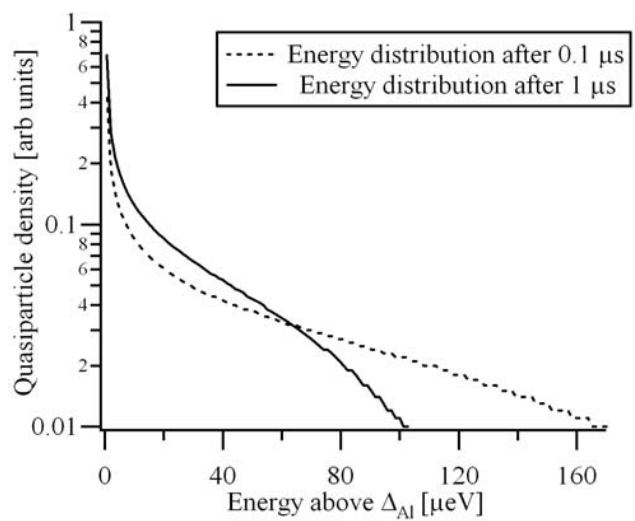

**Fig. 9:**

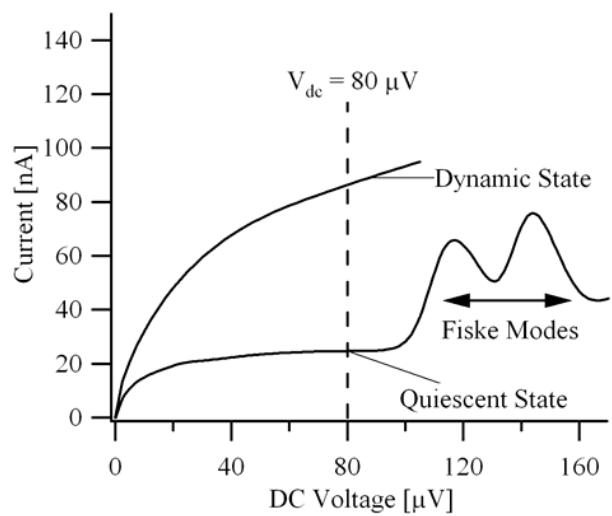

**Fig. 10:**

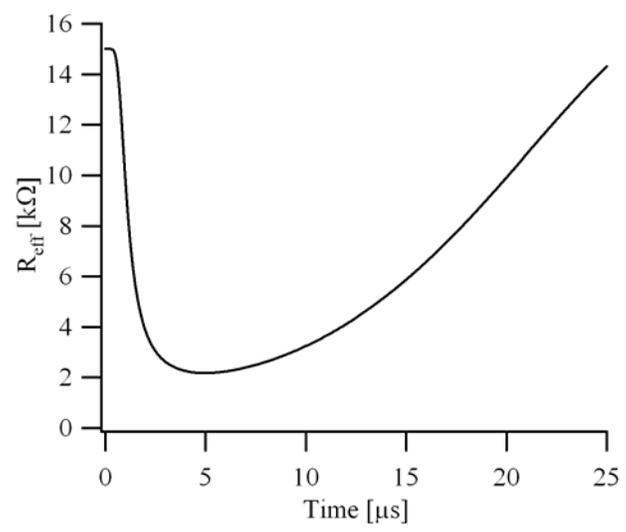

**Fig. 11:**

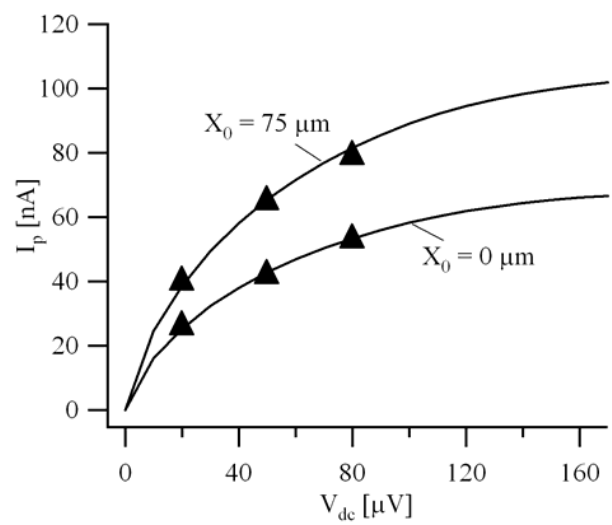

**Fig. 12:**

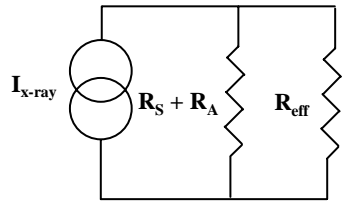 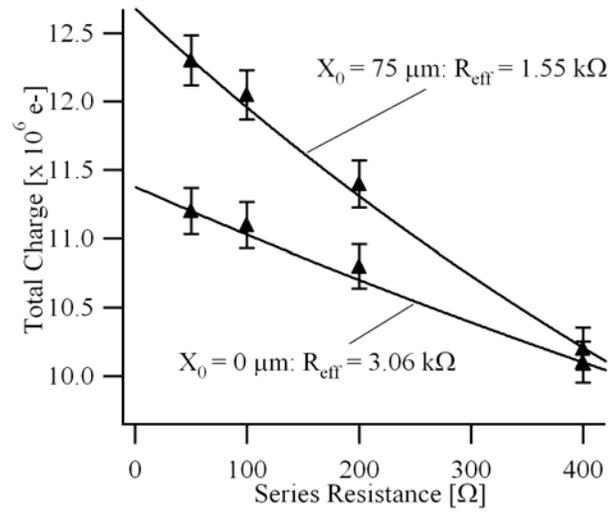

**Fig. 13:**

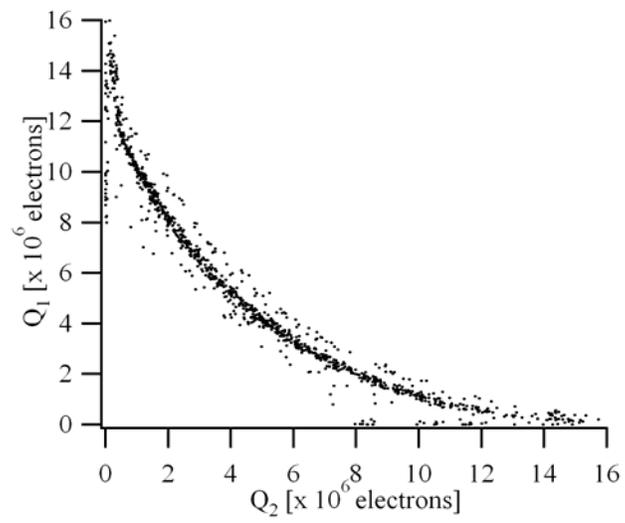

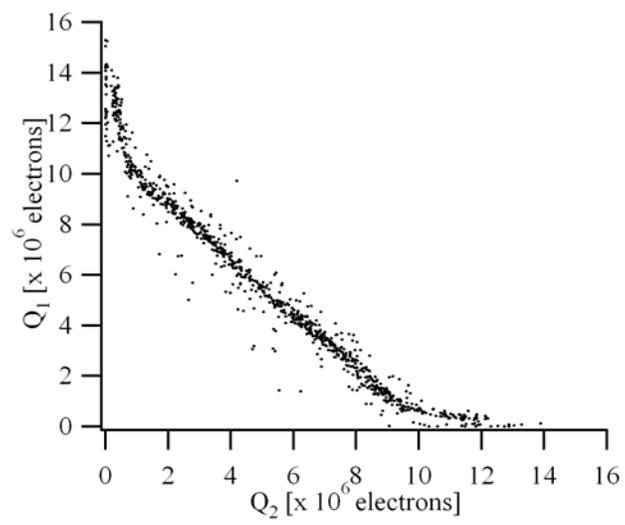

**Fig. 14:**

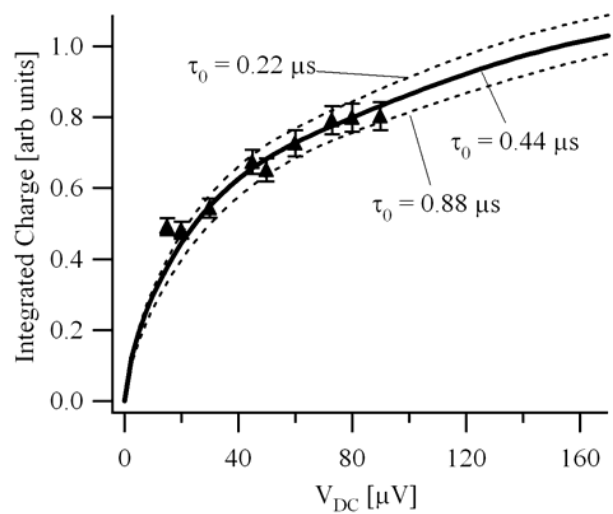

**Fig. 15:**

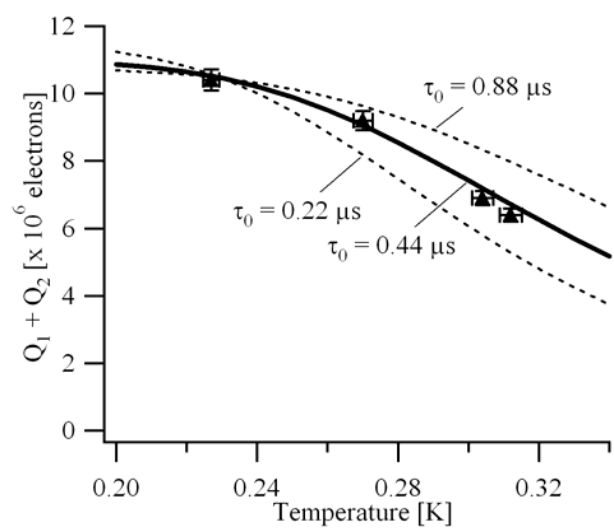